\documentclass[letterpaper]{jpconf}
\usepackage{citesort}

\usepackage{graphicx}
\usepackage{amsmath}
\usepackage{amssymb}
\usepackage{mathtools}
\usepackage{isotope}
\usepackage{dsfont} 

\usepackage{liealg}
\usepackage{wignert}
\usepackage{mcmath}



\newcommand{\bbR}{{\mathbb{R}}}

\newcommand{\St}{{\tilde{S}}}

\newcommand{\Nmax}{{N_\text{max}}}
\newcommand{\Ntot}{{N_\text{tot}}}

\newcommand{\Trel}{{T_\text{rel}}}

\newcommand{\bho}{b_\text{HO}}   
\newcommand{\Ncm}[1][]{{N_\text{c.m.}^{#1}}}  
  
\newcommand{\Omegat}{{\tilde{\Omega}}}

\newcommand{\Ncut}{{N_\text{cut}}}  

\newcommand{\wfgen}{\psi} 
\newcommand{\wfgencm}[1][]{\psi_{\text{c.m.}#1}}
\newcommand{\wfgenin}[1][]{\psi_{\text{in}#1}}

\newcommand{\wfho}{\Psi}

\newcommand{\MeV}{{\mathrm{MeV}}}

\begin{document}


\title{The no-core shell model with general radial bases}

\author{M~A~Caprio$^1$, P~Maris$^2$, J~P~Vary$^2$}
\address{$^1$ Department of Physics, University of Notre Dame, Notre Dame, IN 46556-5670, USA}
\address{$^2$ Department of Physics and Astronomy, Iowa State University, Ames, IA 50011-3160, USA}

\begin{abstract}
Calculations in the \textit{ab initio} no-core shell model (NCSM) have
conventionally been carried out using the harmonic-oscillator
many-body basis. However, the rapid falloff (Gaussian asymptotics) of
the oscillator functions at large radius makes them poorly suited for
the description of the asymptotic properties of the nuclear
wavefunction.  We establish the foundations for carrying out
no-core configuration interaction (NCCI) calculations using a basis built from general
radial functions and discuss some of the considerations which enter into using
such a basis. In particular, we consider the Coulomb-Sturmian basis,
which provides a complete set of functions with a realistic
(exponential) radial falloff.
\end{abstract}



\section{Introduction}
\label{sec-intro}

Significant advances presently are being made towards one of the basic
goals of nuclear theory, namely, an
\textit{ab initio} understanding of the nucleus directly as a system of
interacting protons and neutrons.  Nuclear
interactions motivated by quantum chromodynamics are being developed,
via effective field theory
methods~\cite{entem2003:chiral-nn-potl,epelbaum2009:nuclear-forces},
to provide an underlying Hamiltonian.  It is then
necessary to solve the many-body problem for this Hamiltonian,
obtaining nuclear eigenstates and predictions for observables.  In
no-core configuration interaction (NCCI) approaches such as the
no-core shell model (NCSM)~\cite{navratil2000:12c-ncsm-COMBO}, the
eigenproblem is formulated as a matrix diagonalization problem, in
which the Hamiltonian matrix is represented with respect to a basis of
antisymmetrized products of single-particle states, for the full $A$-body system of nucleons.

Actual NCCI calculations must be carried out in a finite, truncated
space.  
The challenge is to reach a reasonable approximation of the converged
results which would be achieved in the full, untruncated space for
the many-body system.  The success of the calculation is determined by
the rate of convergence of calculated observables (\textit{e.g.}, energies, radii, and electromagnetic moments and transition strengths) with
increasing basis size  and the ability to reliably extrapolate these
results to the full, untruncated many-body
space~\cite{forssen2008:ncsm-sequences,maris2009:ncfc}.  Convergence rates are
sensitive to the choice of single-particle states from
which the many-body basis is constructed, as well as the
truncation scheme used for the many-body basis.

In practice, such calculations have been based almost exclusively on a
harmonic oscillator basis.  
It is worth
recalling the special characteristics of this basis which make it
particularly convenient for use in the nuclear many-body problem:

(1) An exact factorization of center-of-mass and intrinsic
wavefunctions is obtained in many-body calculations when the
oscillator basis is used in conjunction with the $\Nmax$ truncation
scheme, which is based on the total number of oscillator quanta.
Thus, precise removal
of, or correction for, spurious center-of-mass contributions to the
dynamics is possible.

(2) Matrix elements of the nucleon-nucleon two-body interaction are
naturally formulated in the relative oscillator basis.  These matrix elements can easily
be transformed to the two-body oscillator basis, of functions
$\wfho_{n_1l_1}(\vec{r}_1)\wfho_{n_2l_2}(\vec{r}_2)$, by the Moshinsky
transformation~\cite{moshinsky1996:oscillator}.  The simplicity of this
transformation is lost with any other single-particle basis.  

(3) The oscillator functions constitute a complete \textit{discrete}
basis for square-integrable functions.  Many alternative bases, such
as the eigenfunctions of the Schr\"odinger equation for a
\textit{finite-depth} potential such as the Woods-Saxon potential, do
not provide this convenience, since for these bases a continuum of
unbound single-particle states is needed for completeness.

However, there are also significant motivations for moving beyond the
oscillator basis for the nuclear many-body
problem~\cite{stoitsov1998:tho-basis}.  The classic physical
limitation of the oscillator basis, for application to the nuclear
problem, lies in the Gaussian falloff ($\propto e^{-\alpha r^2}$) at
large distance $r$.  In contrast, for particles bound by a
finite-range force, the actual asymptotics are exponential ($\propto
e^{-\beta r}$).  This mismatch in asymptotics, \textit{i.e.}, the
wavefunction tails, between the expansion basis and the physical
system imposes a serious handicap on the convergence of calculations
with increasing basis size.  The
problem is especially significant for observables, such as the
root-mean-square radius or $E2$ strengths, which are sensitive to the
large-$r$ properties of the nuclear wavefunctions.

Here we consider an alternative basis for NCCI calculations, built from Coulomb-Sturmian
functions~\cite{rotenberg1962:sturmian-scatt,weniger1985:fourier-plane-wave}.\footnote{Since the Coulomb-Sturmian single-particle states arise~\cite{rotenberg1962:sturmian-scatt} as solutions to
a general Sturm-Liouville equation, rather than a Schr\"odinger
equation or Hartree-Fock problem, it may be noted that they do not physically correspond to
``shells'' in the conventional sense.  That is, they are not naturally
interpreted as orbitals for independent-particle motion in some
mean-field potential describing the zeroth-order dynamics of the
system.  Therefore, we will use the more inclusive term
\textit{configuration interaction}, rather than specifically
\textit{shell model}.}
Although we focus on the
Coulomb-Sturmian basis, many of the considerations addressed here are
more broadly applicable to alternative single-particle bases for the
NCCI problem.
First, the procedures and results necessary for using the
Coulomb-Sturmian basis for nuclear many-body calculations are outlined
(section~\ref{sec-cs}).  Then, illustrative calculations for
$\isotope[6]{Li}$  are discussed
(section~\ref{sec-calc}).  An expanded discussion may be found in~\cite{caprio20xx:csbasis}.

\section{Coulomb-Sturmian basis}
\label{sec-cs}

\begin{figure}
\begin{center}
\includegraphics*[width=0.9\hsize]{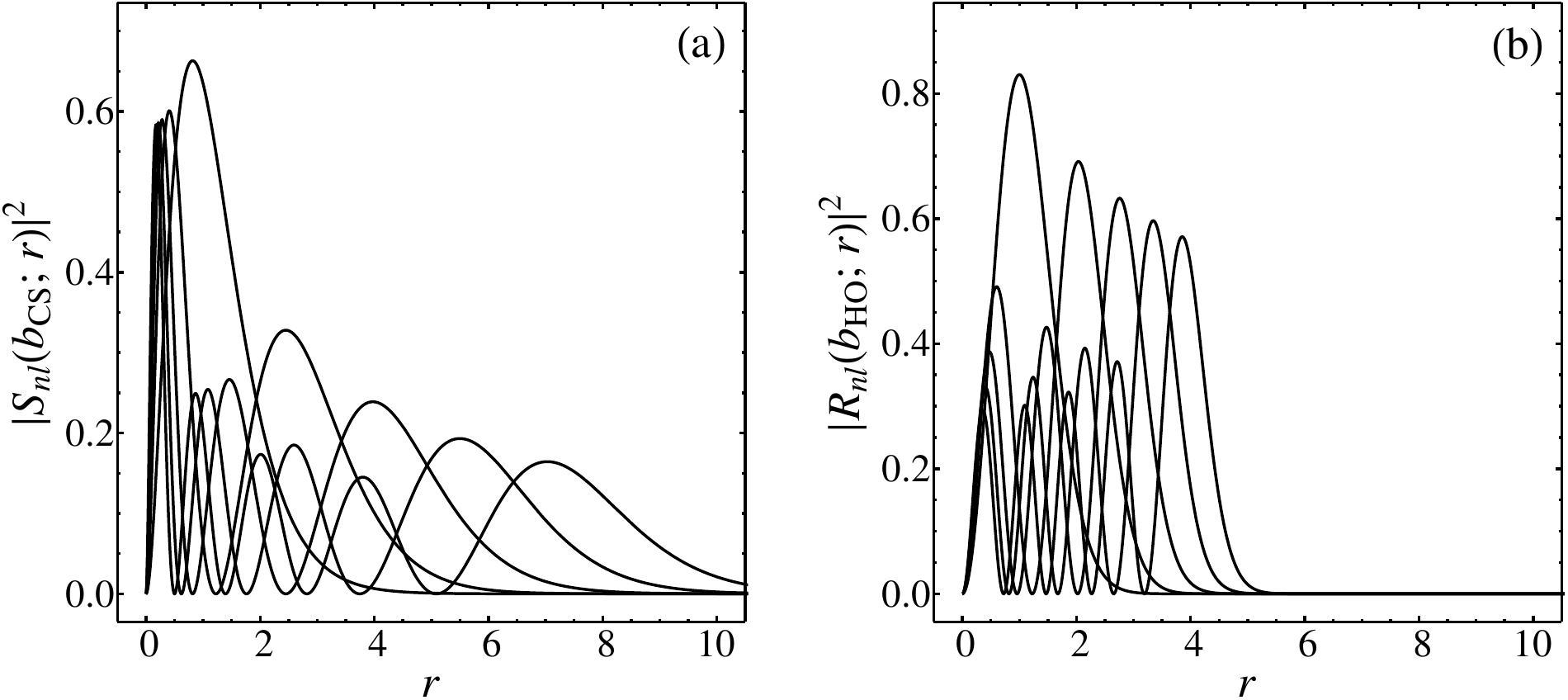}
\end{center}
~\\[-36pt]
\caption{Radial functions for the (a)~Coulomb-Sturmian  and (b)~harmonic oscillator bases, for $l=0$.
The first five functions ($0\leq n \leq 4$) are
shown in each case.
}
\label{fig-functions}
\end{figure}

The
Coulomb-Sturmian
functions~\cite{rotenberg1962:sturmian-scatt,rotenberg1970:sturmian-scatt,weniger1985:fourier-plane-wave},
 have previously been applied to few-body problems in
atomic~\cite{hylleraas1928:helium-sturmian,loewdin1956:natural-orbital,rotenberg1962:sturmian-scatt,rotenberg1970:sturmian-scatt}
and
hadronic~\cite{jacobs1986:heavy-quark-sturmian,keister1997:on-basis,pervin2005:diss}
physics.  They constitute a complete, discrete, orthogonal set of square-integrable
functions, while also possessing realistic exponential asymptotics
appropriate to the nuclear problem.  
The functions on $\bbR^3$ are given by
$\Lambda_{nlm}(\vec{r})=S_{nl}(r)Y_{lm}(\uvec{r})/r$,
with radial wave function
\begin{equation}
\label{eqn-cs-S}
S_{nl}(b;r)\propto(2r/b)^{l+1}
L_n^{2l+2}(2r/b)
e^{-r/b},
\end{equation}
where the $L_n^\alpha$ are generalized Laguerre polynomials, the
$Y_{lm}$ are spherical harmonics, $n$ is the radial or node quantum number,
$l$ and $m$ are the orbital angular momentum and its $z$-projection, and
$b$ is a radial scale parameter, analogous to the oscillator length
parameter for the harmonic oscillator functions.  
The first several Coulomb-Sturmian radial functions, for $l=0$, are
plotted in figure~\ref{fig-functions}(a).  The radial functions for
the harmonic oscillator  are shown for comparison in
figure~\ref{fig-functions}(b).  

Many-body calculations with the Coulomb-Sturmian basis can make use of the
existing computational framework for NCCI calculations, but starting
from the two-body matrix elements of the Hamiltonian in this new basis.
The change of basis transformation has the form
\begin{equation}
\label{eqn-tbme-xform}
\tme{\bar{c}\bar{d};J}{V}{\bar{a}\bar{b};J}=
\sum_{abcd} \toverlap{a}{\bar{a}}\toverlap{b}{\bar{b}}
\toverlap{c}{\bar{c}}\toverlap{d}{\bar{d}}
\, \tme{cd;J}{V}{ab;J},
\end{equation}
where in this expression we label
single-particle orbitals for the oscillator basis by unbarred symbols $a=(n_al_aj_a)$,
$b=(n_bl_bj_b)$, \textit{etc.}, and those for the Coulomb-Sturmian basis
by barred symbols $\bar{a}=(\bar n_a
\bar l_a \bar j_a)$, $\bar{b}=(\bar n_b \bar l_b \bar j_b)$,
\textit{etc.}  The coefficients $\toverlap{a}{\bar{a}}$ required for
the tranformation are given by the overlaps of the harmonic oscillator
and Coulomb-Sturmian radial
wavefunctions, that is, $\toverlap{a}{\bar{a}}=\toverlap{R_{n_al_a}}{S_{\bar{n}_a\bar{l}_a}}
\delta_{(l_a j_a)(\bar l_a \bar j_a)}$, where
$\toverlap{R_{nl}}{S_{\bar{n}l}} 
= \int_0^\infty dr\,
R_{nl}(\bho;r)
S_{\bar{n}l}(b_l;r)$.  The
length parameter for Coulomb-Sturmian and oscillator functions may be
chosen independently in this transformation.\footnote{Moreover, it is of
practical importance to note that orthogonality  of Coulomb-Sturmian wavefunctions
$\Lambda_{nlm}(\vec{r})$ with
\textit{different} $l$ quantum numbers is enforced by the
$Y_{lm}(\uvec{r})$ factor, regardless of the radial wavefunction.
Therefore, the choice of length parameter may be made
independently for each $l$-space, as $b_l$, and orthogonality of the basis of
single-particle states will still be preserved.  The considerations
which enter the selection of the $b_l$ values and the prescription used here are discussed further in~\cite{caprio20xx:csbasis}.}

For actual calculation of the transformed matrix elements, the
infinite sums over single-particle states appearing in the transformation~(\ref{eqn-tbme-xform})
must be
truncated, limited in practice by the available set of
oscillator-basis matrix elements.  In a single-particle space
truncated by number of oscillator quanta, to $N\leq\Ncut$, the sum appearing
in~(\ref{eqn-tbme-xform}) becomes
$\sum_{abcd}^{N_a,N_b,N_c,N_d\leq\Ncut}$.  
The question is whether or not we can
practically include enough oscillator shells in the sum to achieve
sufficient accuracy for the many-body calculation.  Notice that the
comparatively longer tails of the Coulomb-Sturmian functions
[figure~\ref{fig-functions}(a)] imply that they will require
oscillator functions of significantly higher $n$
[figure~\ref{fig-functions}(b)] for their expansion.  One
might therefore ask now what we have gained by changing to an alternative
radial basis, if we must still expand the new basis in terms of high-$n$
oscillator basis functions.  The essential point is that we need only carry out the
expansion at the \textit{two-body} level, where it is still tractable,
to encode the information needed for treatment of the
\textit{$A$-body} problem in the new basis.

Although the truncated transformation~(\ref{eqn-tbme-xform}) is found
to suffice for the matrix elements of the nucleon-nucleon interaction,
with $\Ncut\lesssim13$ (section~\ref{sec-calc}), the matrix elements
of the kinetic energy are significantly more sensitive to the shell
cutoff in the transformation.  These matrix elements may instead be
evaluated directly, making use of the particularly simple form of the
kinetic energy operator.  The NCSM intrinsic Hamiltonian has the form
$H=\Trel + V_{NN}$, where $\Trel$ is the Galilean-invariant relative
kinetic energy operator
\begin{equation}
\Trel\equiv\frac{1}{4Am_N} \sumprime_{ij} (\vec{p}_i-\vec{p}_j)^2\\
=\frac{1}{2Am_N}\Biggl[(A-1) 
\underbrace{\sum_i \vec{p}_i^2}_{\text{one-body}} 
- \underbrace{\sumprime_{ij} \vec{p}_i\cdot\vec{p}_j}_{\begin{subarray}\text{{separable}}\\\text{two-body}\end{subarray}}
\Biggr].
\end{equation}
Here the prime on the summation ${\tsumprime_{ij}}$ indicates $i\neq j$.
This operator is seen to decompose into a one-body part and a
separable two-body part.  The
Coulomb-Sturmian functions have a momentum-space
representation  $\St_{nl}(b;k)$, where $p=\hbar k$, which is given simply in terms of a Jacobi
polynomial~\cite{weniger1985:fourier-plane-wave,keister1997:on-basis}.
Matrix elements of the one-body term are readily
evaluated as radial integrals of $k^2$.  Those of the two-body matrix
elements factorize by Racah's reduction formula~\cite{edmonds1960:am}, as 
\begin{equation}
\tme{cd;J}{\vec{p}_1\cdot\vec{p}_2}{ab;J}=(-)^{j_d+j_a+J}\smallsixj{j_c}{j_d}{J}{j_b}{j_a}{1}
\trme{c}{\vec{p}}{a}\trme{d}{\vec{p}}{b},
\end{equation}
where each factor may again be evaluated in terms of a radial
integral, more specifically, as
$\trme{b}{\vec{p}}{a}\propto
[
\int_0^\infty dk\,\St_{n_bl_b}(b_{l_b};k) \, k \,
\St_{n_al_a}(b_{l_a};k)
]
\trme{l_bj_b}{Y_1}{l_aj_a}$.

\section{\boldmath Calculations for $\isotope[6]{Li}$}
\label{sec-calc}

As a basic illustration of the use of the Coulomb-Sturmian basis for
NCCI calculations, we now consider the nucleus $\isotope[6]{Li}$.  The
code
MFDn~\cite{sternberg2008:ncsm-mfdn-sc08-WORKAROUND,maris2010:ncsm-mfdn-iccs10}
is used for the many-body calculations, taking as its input
Hamiltonian two-body matrix elements obtained as outlined in
section~\ref{sec-cs}.  The calculations are carried out for the
JISP16 interaction~\cite{shirokov2007:nn-jisp16}.
\begin{figure}
\begin{center}
\includegraphics*[width=0.80\hsize]{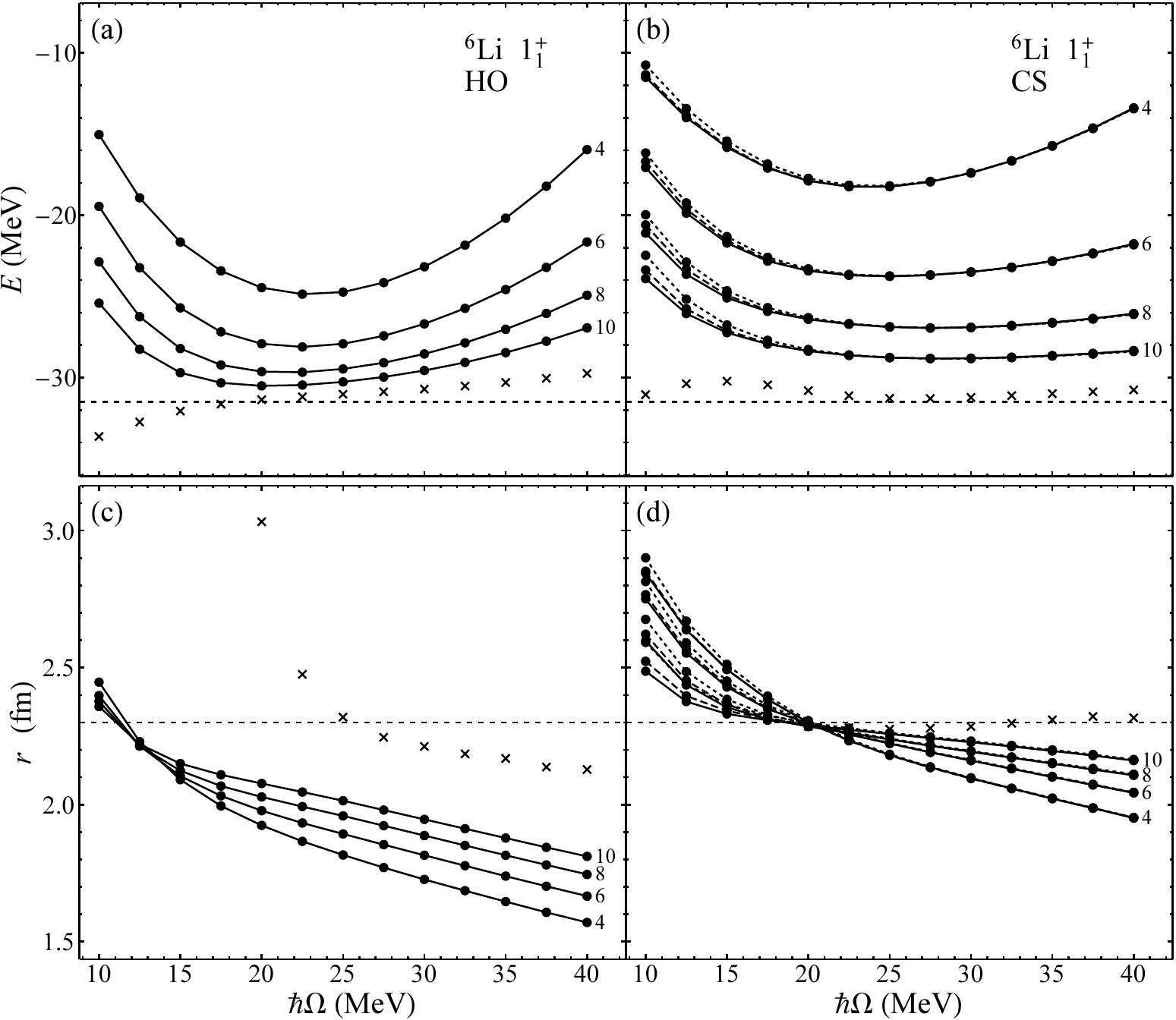}
\end{center}
~\\[-36pt]
\caption{The $\isotope[6]{Li}$ $1^+$ ground
state energy, calculated using the conventional harmonic
oscillator basis~(left) and the Coulomb-Sturmian basis~(right).  Calculated energies~(top) and RMS radii~(bottom) are
plotted as a function of the basis $\hbar
\Omega$ parameter, for $\Nmax=4$, $6$, $8$, and $10$ (successive
curves, as labeled).  For the Coulomb-Sturmian basis, calculations are
shown variously for truncations $\Ncut=9$, $11$, and $13$ (dotted,
dashed, and solid curves, respectively).  Exponentially
extrapolated values are indicated by crosses.
The best extrapolated values from the large-basis calculations of~\cite{cockrell2012:li-ncfc} are shown as horizontal dashed lines.
}
\label{fig-conv-e-r}
\end{figure}

The $\Nmax$ truncation for the harmonic oscillator many-body basis is
defined by the condition $\Ntot \leq N_0+\Nmax$, where 
$\Ntot=\sum_i N_i =\sum_i (2n_i+l_i)$ is the total number of
oscillator quanta, and $N_0$ is the minimal number of oscillator quanta for the given
number of protons and neutrons.  The calculated results depend both on the oscillator length parameter
$b$~--- which is commonly quoted in terms of the oscillator energy
$\hbar\Omega$, where $b=[\hbar/(m_N\Omega)]^{1/2}$, where $m_N$ is the
nucleon mass~--- and the
truncation $\Nmax$ for the basis. 
A natural question is how to truncate the Coulomb-Sturmian many-body basis, where a total number of oscillator quanta is not defined.
However, we
can \textit{formally} carry over the $\Nmax$ truncation scheme, if, for each Coulomb-Sturmian
single-particle state, we simply define $N=2n+l$ and, for each many-body
state, we again define  $\Ntot=\sum_i N_i =\sum_i (2n_i+l_i)$. This choice of many-body truncation
is certainly not unique, but it provides a reasonable starting point
for further exploration, and it facilitiates comparison of convergence
rates obtained using the oscillator and Coulomb-Sturmian bases, since
the dimensions of the many-body spaces are then the same in both cases.

First, let us review the conventional oscillator results for the
energy  of the $1^+$ ground state, shown in
figure~\ref{fig-conv-e-r}(a).  
For any
given choice of $\Nmax$, the energy has a variational minimum for some value
of $\hbar\Omega$, $\sim20\,\MeV$ in this example.  With increasing
$\Nmax$, the ground-state eigenvalue approaches that for  the infinite, untruncated
space.  The horizontal dashed line in
figure~\ref{fig-conv-e-r} indicates the best extrapolation from prior
calculuations in an $\Nmax=16$ space~\cite{cockrell2012:li-ncfc}.

Calculations of the ground-state energy using the Coulomb-Sturmian
basis are shown in figure~\ref{fig-conv-e-r}(b).  At each $\Nmax$ the
variational minimum energy obtained with the Coulomb-Sturmian basis in
figure~\ref{fig-conv-e-r}(b) is substantially higher than that
obtained with the oscillator basis in figure~\ref{fig-conv-e-r}(a).
However, the energies obtained with the Coulomb-Sturmian basis are
also falling significantly more rapidly with increasing $\Nmax$, and
the quality of extrapolation appears to be comparable.  We note that
these exploratory results have not yet probed several variational
freedoms available with the Coulomb-Sturmian basis, both in the length
parameters $b_l$ (see~\cite{caprio20xx:csbasis}) and in the many-body
truncation scheme.  The robustness of the
many-body calculations against the truncation $\Ncut$ in the
change-of-basis transformation~(\ref{eqn-tbme-xform}) may also be verified from
figure~\ref{fig-conv-e-r}, where calculations based on two-body matrix
elements obtained with $\Ncut=9$, $11$, and $13$ are overlaid.  The
results for the ground state energy for $\hbar\Omega\gtrsim20\,\MeV$
are highly stable with respect to this cutoff. This range safely
covers the variational minimum.  If necessary, it would also be
practicable to carry out transformations in larger oscillator spaces.

The root-mean-square (RMS) radius presents challenges for convergence
in NCCI calculations with the conventional oscillator
basis~\cite{bogner2008:ncsm-converg-2N}, since it is more specifically sensitive to
the large-$r$ asymptotic properties of the nuclear wavefunction.  We compare the results
obtained in the oscillator basis, shown in figure~\ref{fig-conv-e-r}(c),
with those obtained in the Coulomb-Sturmian basis, shown in
figure~\ref{fig-conv-e-r}(d).
The extrapolated values obtained for
$\hbar\Omega\gtrsim20\,\MeV$, \textit{i.e.}, above the crossover
point, are reasonably insensitive to $\hbar\Omega$ and are consistent
with the best estimate from large oscillator-basis calculations~\cite{cockrell2012:li-ncfc}.  
It would appear that the rate of convergence of the RMS radius
obtained with the Coulomb-Sturmian basis is superior to that obtained
with the conventional oscillator basis.  However, further systematic
investigation is required, especially into the stability of
extrapolations with increasing $\Nmax$, before general conclusions may
be drawn.

The dominant concern in using any basis other than the
harmonic oscillator basis, with $\Nmax$ truncation, is incomplete separation of
center-of-mass and intrinsic dynamics.  There are several aspects to
consider: the degree of separation arising in calculations
using the Coulomb-Sturmian basis, the spurious state spectrum obtained
in such calculations, and the extent to which a Lawson term~\cite{gloeckner1974:spurious-com} can be
used to purge spurious excitations from the low-lying spectrum.  Here
we briefly discuss only the
first consideration, \textit{i.e.}, separation.  However, it is also found that the spurious states
are well-separated in the spectrum and are amenable to management with a Lawson term (see~\cite{caprio20xx:csbasis}).
\begin{figure}
\begin{center}
\includegraphics*[width=0.65\hsize]{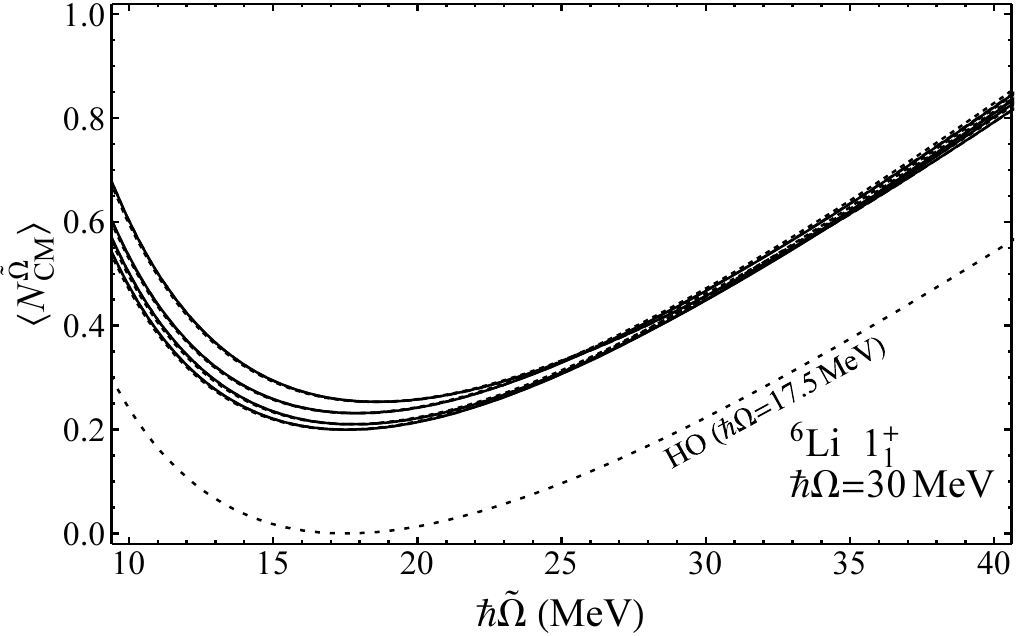}
\end{center}
~\\[-36pt]
\caption{Expectation value of the number operator $\Ncm[\Omegat]$ for
center-of-mass oscillator quanta, as a function of oscillator energy
$\hbar\Omegat$.  These calculations are
for the $\isotope[6]{Li}$ $1^+$ ground
state, using the Coulomb-Sturmian basis, 
with $\hbar\Omega=30\,\mathrm{MeV}$.
Calculations
are shown for $\Nmax=4$, $6$, $8$, and $10$ (successive
curves, top to bottom), and for $\Ncut$ values as indicated in the
caption to figure~\ref{fig-conv-e-r}.  The analogous curve
expected for a pure harmonic oscillator $0s$ function, with
$\hbar\Omega=17.5\,\MeV$, is also
shown (dotted curve, labeled).
}
\label{fig-Ncm-distrib}
\end{figure}

A first indication of the degree of separation in the many-body
eigenstate is provided by the expectation
value of the number operator for center-of-mass harmonic oscillator
quanta, defined for an arbitrary center-of-mass harmonic oscillator energy $\hbar\Omegat$ by
\begin{equation}
\label{eqn-ncm}
\Ncm[\Omegat]
\equiv
\frac{1}{\hbar\Omegat}\biggl(
\frac{P^2}{2Am_N}+\frac{Am_N\Omegat^2R^2}{2}-\frac{3\hbar\Omegat}{2}
\biggr),
\end{equation}
where $\vec{R}$ and $\vec{P}$ are the center-of-mass coordinate and
momentum, respectively.
As noted by Hagen~\textit{et
al.}~\cite{hagen2009:coupled-cluster-com},
\textit{if} separation occurs, as
$\wfgen(\vec{r}_i;\vec{\sigma}_i)=\wfgencm(\vec{R})\wfgenin(\vec{r}_{ij};\vec{\sigma}_i)$,
and \textit{if} $\wfgencm(\vec{R})$ happens to be an oscillator $0s$
function, corresponding to some oscillator energy $\hbar\Omegat$, then
the many-body wavefunction will have $\tbracket{\Ncm[\Omegat]}=0$.  
The expectation value $\tbracket{\Ncm[\Omegat]}$ is shown as a
function of $\hbar\Omegat$ for the $\isotope[6]{Li}$ ground
state in figure~\ref{fig-Ncm-distrib}, for the Coulomb-Sturmian basis
calculation with basis $\hbar\Omega=30\,\mathrm{MeV}$ and no Lawson term.
The minimum
 of $\tbracket{\Ncm[\Omegat]}$ is obtained at
$\hbar\Omegat\approx17.5\,\MeV$.
The minimum value
decreases with increasing $\Nmax$, but it appears to be converging
towards a \textit{nonzero} $\tbracket{\Ncm[\Omegat]}\sim0.2$.
 The fact that
expectation values significantly less than unity are
obtained in the calculations indicates that a $0s$ oscillator function
dominates the center-of-mass motion, and that an approximate
separation of center-of-mass and intrinsic functions is spontaneously
arising.  However, the nonzero limit indicates that, as the full space
is approached, the separated center-of-mass function is \textit{not}
strictly taking the form of a $0s$ oscillator function.

\section{Summary}
\label{sec-concl}

Although the conventional oscillator basis for the NCCI approach provides
definite benefits,
namely, exact center-of-mass factorization in the $\Nmax$ truncation  and the simplicity of the Moshinsky transformation for
generating two-body matrix elements, it also exhibits
nonphysical Gaussian asymptotics at large distances.  The Coulomb-Sturmian
functions form a complete, discrete set of
square-integrable functions with realistic
exponential asymptotics.  A principal goal  in using such a basis
is to obtain improved convergence of observables which are sensitive to the asymptotics,
such as RMS radii and $E2$ matrix elements.    Such a basis  might
be particularly appropriate to halo nuclei, where the mismatch with
the oscillator functions at large distances is particularly severe.  
The computational
framework for the many-body calculation has the standard structure for an
$nlj$ single-particle basis, and the interaction matrix elements are
transformed from the harmonic-oscillator basis, except that relative kinetic
energy matrix elements (and, in fact, those for $\Ncm$ and $r^2$, as
well) are calculated directly in the Coulomb-Sturmian basis.  
The initial
exploratory calculations considered here indicate that convergence
properties are promising.  Moreover, the spurious center-of-mass dynamics are
found to be tractable.


\ack

We thank M.~Pervin and W.~N.~Polyzou for pointing out the relevance of
the Coulomb-Sturmian basis.  This work was supported by the Research
Corporation for Science Advancement through the Cottrell Scholar
program, by the US Department of Energy under Grants
No.~DE-FG02-95ER-40934, DE-FC02-09ER41582 (SciDAC/UNEDF), and
DE-FG02-87ER40371, and by the US National Science Foundation under
Grant No.~0904782. Computational resources were provided by the
National Energy Research Supercomputer Center (NERSC), which is
supported by the U.S. Department of Energy
under Contract No.~DE-AC02-05CH11231.


\section*{References}

\providecommand{\newblock}{}


\end{document}